\documentclass[%
 aip,
rsi,%
 amsmath,amssymb,
reprint,%
]{revtex4-1}

\usepackage{graphicx}
\usepackage{dcolumn}
\usepackage{bm}
\usepackage[mathlines]{lineno}
\usepackage{txfonts} 

\usepackage[
pdfauthor={},
pdftitle={},
bookmarks=false,
bookmarksnumbered,
pdftex,
pdfstartview=FitH 
]{hyperref}
\hypersetup{colorlinks=true, 
linkcolor=blue,
urlcolor=blue,
citecolor=blue}
\usepackage[all]{hypcap} 

\begin{document}

\title{Method for Transferring High-Mobility CVD-Grown Graphene with Perfluoropolymers}

\author{Jianan Li}
\affiliation{%
Department of Physics and Astronomy, University of Pittsburgh, Pittsburgh,  Pennsylvania 15260, USA
}%

\author{Jen-Feng Hsu}
\affiliation{%
Department of Physics and Astronomy, University of Pittsburgh, Pittsburgh,  Pennsylvania 15260, USA
}%
\author{Hyungwoo Lee}
\affiliation{%
Department of Materials Science and Engineering, University of Wisconsin-Madison, Madison, Wisconsin 53706, USA
}%

\author{Shivendra Tripathi}
\affiliation{%
Department of Physics and Astronomy, University of Pittsburgh, Pittsburgh,  Pennsylvania 15260, USA
}%
\author{Qing Guo}
\affiliation{%
Department of Physics and Astronomy, University of Pittsburgh, Pittsburgh,  Pennsylvania 15260, USA
}%
\author{Lu Chen}
\affiliation{%
Department of Physics and Astronomy, University of Pittsburgh, Pittsburgh,  Pennsylvania 15260, USA
}%
\author{Mengchen Huang}
\affiliation{%
Department of Physics and Astronomy, University of Pittsburgh, Pittsburgh,  Pennsylvania 15260, USA
}%
\author{Shonali Dhingra}
\affiliation{%
Department of Physics and Astronomy, University of Pittsburgh, Pittsburgh,  Pennsylvania 15260, USA
}%
\author{Jung-Woo Lee}
\affiliation{%
Department of Materials Science and Engineering, University of Wisconsin-Madison, Madison, Wisconsin 53706, USA
}%
\author{Chang-Beom Eom}
\affiliation{%
Department of Materials Science and Engineering, University of Wisconsin-Madison, Madison, Wisconsin 53706, USA
}%
\author{Patrick Irvin}
\affiliation{%
Department of Physics and Astronomy, University of Pittsburgh, Pittsburgh,  Pennsylvania 15260, USA
}%
\author{Jeremy Levy}
\affiliation{%
Department of Physics and Astronomy, University of Pittsburgh, Pittsburgh,  Pennsylvania 15260, USA
}%
\author{Brian D'Urso}
\email[Author to whom correspondence should be addressed. Electronic mail: ]{dursobr@pitt.edu}
\affiliation{%
Department of Physics and Astronomy, University of Pittsburgh, Pittsburgh,  Pennsylvania 15260, USA
}%

\date{\today}

\begin{abstract}

The transfer of graphene grown by chemical vapor deposition (CVD) using amorphous polymers represents a widely implemented method for graphene-based electronic device fabrication. However, the most commonly used polymer, poly(methyl methacrylate) (PMMA), leaves a residue on the graphene that limits the mobility. Here we report a method for graphene transfer and patterning that employs a perfluoropolymer---Hyflon---as a transfer handle and to protect graphene against contamination from photoresists or other polymers. CVD-grown graphene transferred this way onto LaAlO$_3$/SrTiO$_3$ heterostructures is atomically clean, with high mobility (\~{}30,000 cm$^2$V$^{-1}$s$^{-1}$) near the Dirac point at 2 K and clear, quantized Hall and magneto-resistance. Local control of the LaAlO$_3$/SrTiO$_3$ interfacial metal-insulator transition---through the graphene---is preserved with this transfer method. The use of perfluoropolymers such as Hyflon with CVD-grown graphene and other 2D materials can readily be implemented with other polymers or photoresists.

\end{abstract}

\maketitle

Single-layer graphene has proven to be an extraordinary 2D material system due to its unique properties such as a high mobility 2D electron gas and Dirac behavior of electrons\cite{Novoselov_Electric_2004,Novoselov_Two_2005,Geim_The_2007,Geim_Graphene_2009}. A key factor influencing the high mobility of graphene-based electronic devices is the method for fabricating and transferring graphene onto a given substrate. Current state-of-the-art high-mobility graphene devices are created from mechanical exfoliation followed by encapsulation with hexagonal boron nitride (h-BN)\cite{Dean_Boron_2010}. For applications requiring an arbitrary substrate and graphene shape, CVD growth followed by transfer of graphene on a PMMA scaffold is preferred \cite{Li_Transfer_2009,Reina_Transferring_2008,Reina_Large_2009}. However, residual PMMA remaining after graphene transfer is known to be a source of electron scattering, significantly limiting the graphene mobility\cite{Pirkle_The_2011,Lin_Graphene_2012,Cheng_Toward_2011}. Annealing in H$_2$/Ar environment can partially remove the PMMA residue, but this process can introduce structural defects\cite{Lin_Graphene_2012} or greatly increase coupling to the substrate, resulting in extrinsic doping and even deterioration of mobility\cite{Cheng_Toward_2011}.

We have developed a graphene transfer method that incorporates an amorphous perfluoropolymer, Hyflon\textsuperscript{TM} AD 60 (Solvay), to physically separate a hard-to-remove polymer from the graphene. Hyflon has been widely used in membrane applications such as fuel cells\cite{Arcella_Hyflon_2005,Merlo_Membrane_2007,Zhang_Recent_2012} due to the chemical inertness of C-F bond and high selectivity to solvents, such as Fluorinert\textsuperscript{TM} FC-40. Recently, it has been reported that inserting a Hyflon membrane between graphene and a hydrophilic substrate like SiO$_2$ can reduce the extrinsic p-type doping in graphene by preventing water vapor adsorption to the dangling bonds on the substrate\cite{Mattevi_Solution_2012}. 

The substrate we chose is a heterostructure consisting of bulk TiO$_2$-terminated SrTiO$_3$ (STO) with a thin layer of LaAlO$_3$ (LAO) on top. The LAO is grown with pulsed laser deposition using methods described elsewhere\cite{Huang_Electric_2015}. The LAO/STO system is known for the formation of a 2D electron gas (2DEG) at the interface when the thickness of LAO is larger than a 4 unit cell critical thickness\cite{Ohtomo_A_2004,Thiel_Tunable_2006}. Many other interesting properties have been reported on the LAO/STO interface as well, such as magnetism \cite{Brinkman_Magnetic_2007,Li_Coexistence_2011,Bi_Room_2014}, superconductivity\cite{Reyren_Superconducting_2007}, and electron pairing without superconductivity\cite{Cheng_Electron_2015}. For the LAO/STO samples with a critical thickness (3.4 unit cells) the interface is nominally insulating; however, an insulator-to-metal transition can be achieved by scanning the surface with a positively biased conductive atomic force microscope (c-AFM) tip\cite{Cen_Nanoscale_2008,Cen_Oxide_2009}. The graphene/LAO/STO system has been investigated previously using mechanically exfoliated graphene; it was found that c-AFM writing of LAO/STO nanostructures can be performed in the presence of a graphene top layer\cite{Huang_Electric_2015}. We have fabricated graphene/LAO/STO heterostructure with CVD-grown graphene and and transferred using a PMMA scaffold\cite{jnawali2016room}, but the residue of PMMA on LAO and graphene was difficult to remove. 

\begin{figure*}[tp]
\includegraphics[width=0.8\textwidth]{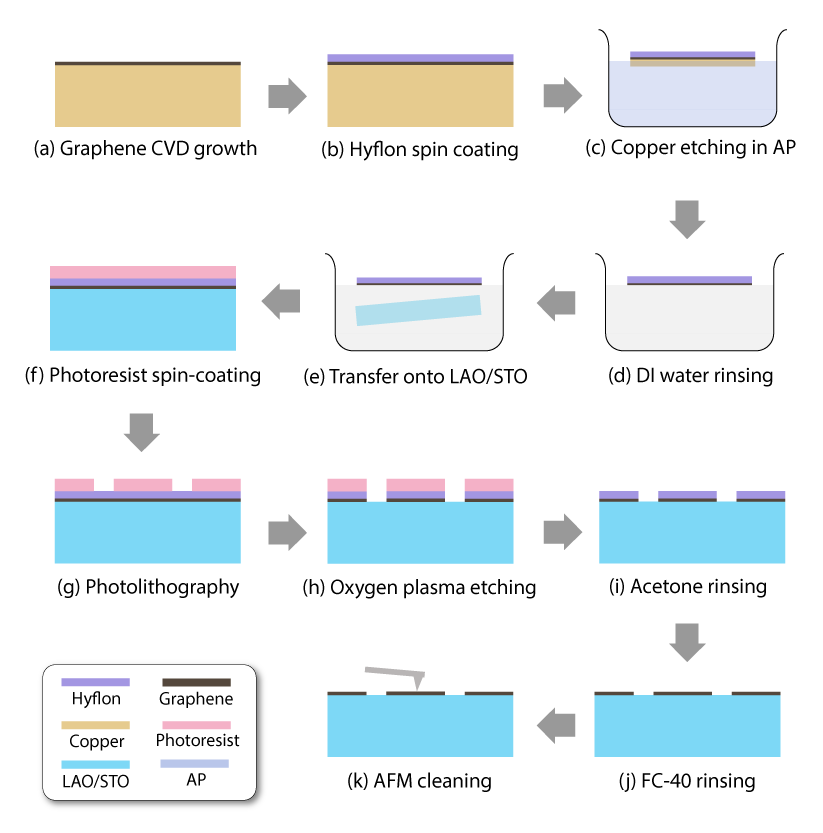}
\label{FIG1}
\caption{Flow chart for graphene growth, transferring and patterning. (a) Graphene is grown on ultra-flat copper surface with atmosphere pressure CVD. (b) Hyflon is spin-coated on top of the graphene after growth. (c) The copper substrate is etched with ammonium persulfate (AP) and (d) subsequently rinsed in DI water. (e) Hyflon/graphene on the DI water surface is transferred onto the substrate by lifting the substrate in the water. Then the sample is soft-baked. (f) Photoresist is spin-coated on top of the Hyflon. (g) The photoresist is patterned using standard photolithography. (h) Hyflon and graphene on the patterned area are etched using oxygen plasma. (i) Photoresist on the un-patterned area is rinsed with acetone and IPA. (j) Hyflon is rinsed with FC-40. (k) Hyflon residue on the sample is cleaned with AFM contact scanning.}
\end{figure*}

The sample preparation procedures are illustrated in FIG. \ref{FIG1}. Graphene is grown on ultra-flat diamond turned copper substrates using atmospheric pressure chemical vapor deposition (APCVD)\cite{Dhingra_Chemical_2014} (FIG. \ref{FIG1}(a)). The size of single crystal graphene domains can reach 70 $\mathrm{\muup m}$. 0.5\% Hyflon FC-40 solution is spin-coated and baked on a 95 $^\circ$C hotplate for 1 minute (FIG. \ref{FIG1}(b)). The typical Hyflon thickness is 80 nm. The Hyflon-coated graphene/Cu is placed on the surface of 1 mol/L ammonium persulfate (AP) solution for 3 hours until the copper is completely dissolved (FIG. \ref{FIG1}(c)), and then the Hyflon with graphene is captured and rinsed in DI water three times (FIG. \ref{FIG1}(d)). The Hyflon with graphene is left floating on the water surface. An LAO/STO substrate with pre-patterned electrodes is immersed underneath the graphene and then lifted to the water surface so that the Hyflon with graphene rests atop the LAO/STO surface (FIG. \ref{FIG1}(e)). The LAO/STO with Hyflon and graphene is baked at 50 $^\circ$C in an oven for 3 minutes. AZ4210 photoresist is spin-coated onto the Hyflon (FIG. \ref{FIG1}(f)) and patterned with 365 nm UV exposure (FIG. \ref{FIG1}(g)). The Hyflon and graphene in the patterned region is etched by oxygen plasma (FIG. \ref{FIG1}(h)). The undeveloped photoresist is removed with acetone and isopropanol (IPA) (FIG. \ref{FIG1}(i)), leaving only Hyflon on the patterned graphene. The sample is placed into hot FC-40 liquid (180 $^\circ$C), and then shaken for 48 hours while the liquid cools slowly to room temperature (FIG. \ref{FIG1}(j)). Finally, the residual Hyflon is cleared by AFM scanning.

\begin{figure*}[tp]
\includegraphics[width=0.8\textwidth]{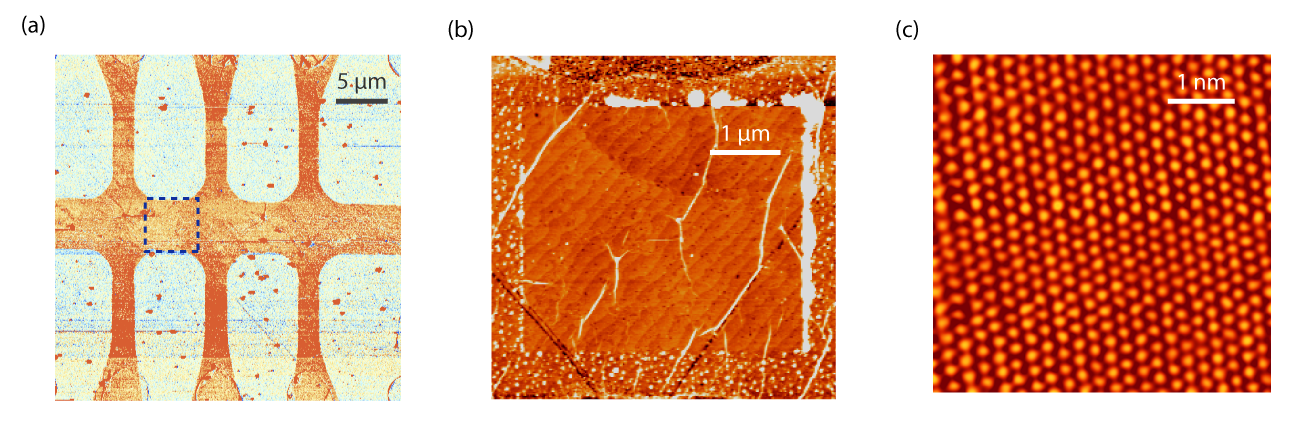}
\label{FIG2}
\caption{(a) AFM AC scan phase image of the as-patterned graphene sample. Orange region shows the hall-bar geometry of graphene on LAO/STO substrate. Fine grains of Hyflon residue particles can be seen on the graphene while larger particles are visible over the entire region. (b) Close-up AC mode height image of the squared region in (a) after AFM contact scan. Hyflon residue accumulates on the boundaries of the square region of contact mode scan. Unit cell terraces of  LAO under the graphene were visible within the cleaned square region. Outside the square particles of Hyflon residue are visible. (c) Room temperature STM image of the graphene shows that the graphene is atomically clean.}
\end{figure*}

Patterned graphene is distinguishable from bare LAO on the phase channel image from AFM AC mode scan (FIG. \ref{FIG2}(a)).   Hyflon residue is still visible on the graphene right after Hyflon removal with FC-40, with a typical thickness of 1 nm. Larger particles of several hundreds of nanometers size from the FC-40 are scattered over the entire sample. The thin layer of Hyflon and larger particles can be moved around on the surface using contact-mode AFM scanning (FIG. \ref{FIG2}(b)). Within the contact scan area, 4 {\AA} unit-cell steps of LAO are distinguishable, in addition to some wrinkles on the graphene. Residue from the Hyflon accumulates at the perimeter of the contact-scanned area. In some places where the graphene has fractured due to prior contact-mode AFM scanning, the graphene can peel from the LAO surface (not shown in the figure).  Generally, contact-mode AFM scanning leaves the graphene surface atomically flat and clean. FIG. \ref{FIG2}(c) shows a room-temperature STM image of the graphene/LAO/STO surface after AFM cleaning has been performed.

\begin{figure*}[tp]
\includegraphics[width=0.8\textwidth]{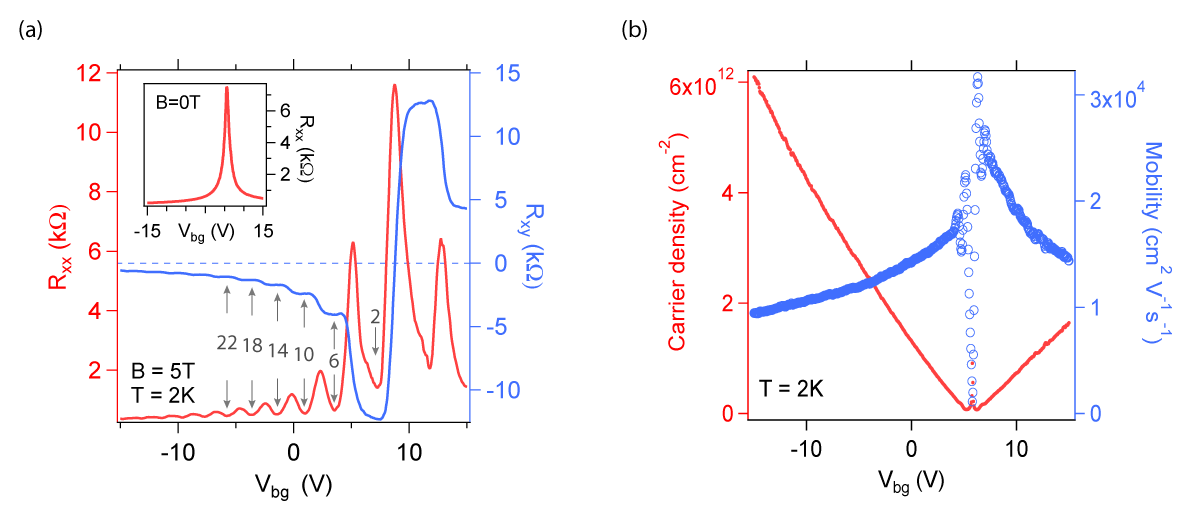}
\label{FIG3}
\caption{Transport measurement on the graphene. (a) Quantized magnetoresistance  (red) and Hall resistance $R_{xy}$ (blue) of the patterned graphene as a function backgate voltage at fixed magnetic field B = 5 T. Well-defined plateaus corresponding to the quantum Hall states $h/\nu e^2$ in $R_{xy}$ with vanishing $R_{xx}$ could be seen, with filling factor $\nu = 4(N+1/2)$ (indicated by the grey numbers with arrows). Inset: magnetoresistance as a function of backgate voltage at B = 0 T. (b) Mobility and carrier density as a function of backgate voltage. The graphene mobility was close to 10,000 cm$^2$V$^{-1}$s$^{-1}$ away from the Dirac point and 30,000 cm$^2$V$^{-1}$s$^{-1}$ near the Dirac point. T = 2 K.}
\end{figure*}

The low-temperature ($T = 2 \mathrm{~K}$) graphene mobility increases two-fold after cleaning, consistent with other reports\cite{Goossens_Mechanical_2012,Lindvall_Cleaning_2012}. At 2 K, the Dirac point was at +7 V (FIG. \ref{FIG3}(a) inset) meaning that the ungated graphene is slightly p-type doped on LAO/STO. Away from the Dirac point the mobility is about 10,000 cm$^2$V$^{-1}$s$^{-1}$ (FIG. \ref{FIG3}(b)), while it reaches 30,000 cm$^2$V$^{-1}$s$^{-1}$ close to the Dirac point, which is higher than most of the wet transferred CVD graphene reported. The hole carrier density reaches 6$\times$10$^{12}$ cm$^{-2}$ at a backgate voltage of only -15 V, owing to the high dielectric constant of STO at low temperatures\cite{Weaver_Dielectric_1959}. On the electron side however, the carrier density does not exceed 1.5$\times 10^{12}$ cm$^{-2}$ (not shown in the figure), likely due to the formation of a 2D electric gas between the sub-critical thickness LAO and STO at positive backgates\cite{Thiel_Tunable_2006} and subsequent screening of the electric field. When a 5 tesla magnetic field is applied at 2 K, clear quantization of magnetoresistance $R_{xx}$ and Hall resistance $R_{xy}$ is observed as a function of backgate voltage. The filling factor of the quantum Hall states $h/\nu e^2$ satisfy the known single-layer graphene quantization condition $\nu = 4(N+1/2)$, as  indicated in FIG. \ref{FIG3}(a). 

\begin{figure*}[tp]
\includegraphics[width=0.8\textwidth]{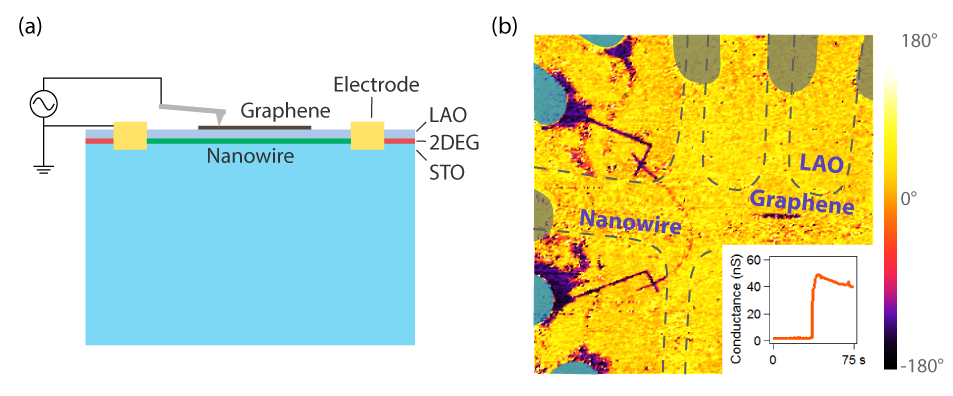}
\label{FIG4}
\caption{Piezoresponse measurement of the nanowire on graphene/LAO/STO heterostructure. (a) The nanowire has been written between two grounded interface electrodes. An AC voltage is applied onto the c-AFM tip; a difference in carrier densities between the nanowire and unwritten areas causes a difference in piezoelectric response detected by the c-AFM tip. (b) PFM phase image of a graphene/LAO/STO sample after nanowire lithography on the interface of LAO/STO. The image offset and range are adjusted so that the features of interest are clear. Boundaries of graphene are marked with grey dashed lines. The gray regions are the electrodes in contact with the graphene; the blue regions are the LAO/STO interface electrodes. Funnels and lines with deep purple color are the virtual electrodes and nanowires on the LAO/STO interface written by the c-AFM tip. The section of nanowire under graphene is slightly diminished. Inset: conductance jump between two interface electrode when a nanowire is written under graphene area. Typical conductance is 50 nS.}
\end{figure*}

It has previously been demonstrated that conductive nanostructures can be ``written'' at the LAO/STO in the presence of exfoliated graphene\cite{Huang_Electric_2015}. Here we investigate whether this property is preserved for CVD-grown graphene transferred with the Hyflon-based procedure described above. FIG. \ref{FIG4}(a) illustrates the configuration under which conductive nanostructures are created and probed. The inset of FIG. \ref{FIG4}(b) shows the characteristic current jump that results when a conductive nanostructure is created between two interfacial electrodes. Typical conductance of a nanowire underneath the graphene area is about 50 nS.  In this case, the path of the current lies underneath the graphene Hall bar channel. Piezoforce microscopy (PFM) is employed as a method for determining the local carrier density underneath the graphene. This technique has been previously established for LAO/STO, where it was found that the carrier density is proportional to the piezoelectric response (up to a constant offset)\cite{Huang_Direct_2013}. FIG. \ref{FIG4}(b) shows the nanowire (deep purple areas) written underneath and outside the graphene (enclosed by dashed gray lines). Electrodes connected to the interface and graphene are painted with blue and grey color respectively. Several nanowire writings were attempted before the PFM imaging, therefore some other high carrier density features are visible. The nanowire under the graphene has a slightly smaller piezoelectric response due to electrical shielding by the graphene, causing the contrast to be diminished somewhat. 

In summary, we have developed a robust method for successful transfer of graphene onto the surface of LaAlO$_3$/SrTiO$_3$ using Hyflon. Mobility of the graphene could reach 10,000 to 30,000 cm$^2$V$^{-1}$s$^{-1}$ at 2 K, which is higher than most reports of CVD-grown graphene after wet transfer and patterning with PMMA. Clear quantization of magnetoresistance and Hall resistance plateaus is observed. The method of using Hyflon as a buffer layer against other photoresists or polymer may find more widespread use in the growing field of two-dimensional materials that are synthesized by means other than mechanical exfoliation. For the specific case of graphene/LAO/STO, the combination of high mobility with the capability of locally controlling the electron density at the LAO/STO interface opens new avenues for interesting and potentially fruitful interactions between these two intimately coupled electronic systems.

Support for this research by ONR (N00014-13-1-0806), AFOSR (FA9550-10-1-0524), NSF (DMR-1124131), AFOSR (FA9550-15-1-0334) and AOARD  (FA2386-15-1-4046) is gratefully acknowledged.

\bibliography{main}

\begin{thebibliography}{31}%
\makeatletter
\providecommand \@ifxundefined [1]{%
 \@ifx{#1\undefined}
}%
\providecommand \@ifnum [1]{%
 \ifnum #1\expandafter \@firstoftwo
 \else \expandafter \@secondoftwo
 \fi
}%
\providecommand \@ifx [1]{%
 \ifx #1\expandafter \@firstoftwo
 \else \expandafter \@secondoftwo
 \fi
}%
\providecommand \natexlab [1]{#1}%
\providecommand \enquote  [1]{``#1''}%
\providecommand \bibnamefont  [1]{#1}%
\providecommand \bibfnamefont [1]{#1}%
\providecommand \citenamefont [1]{#1}%
\providecommand \href@noop [0]{\@secondoftwo}%
\providecommand \href [0]{\begingroup \@sanitize@url \@href}%
\providecommand \@href[1]{\@@startlink{#1}\@@href}%
\providecommand \@@href[1]{\endgroup#1\@@endlink}%
\providecommand \@sanitize@url [0]{\catcode `\\12\catcode `\$12\catcode
  `\&12\catcode `\#12\catcode `\^12\catcode `\_12\catcode `\%12\relax}%
\providecommand \@@startlink[1]{}%
\providecommand \@@endlink[0]{}%
\providecommand \url  [0]{\begingroup\@sanitize@url \@url }%
\providecommand \@url [1]{\endgroup\@href {#1}{\urlprefix }}%
\providecommand \urlprefix  [0]{URL }%
\providecommand \Eprint [0]{\href }%
\providecommand \doibase [0]{http://dx.doi.org/}%
\providecommand \selectlanguage [0]{\@gobble}%
\providecommand \bibinfo  [0]{\@secondoftwo}%
\providecommand \bibfield  [0]{\@secondoftwo}%
\providecommand \translation [1]{[#1]}%
\providecommand \BibitemOpen [0]{}%
\providecommand \bibitemStop [0]{}%
\providecommand \bibitemNoStop [0]{.\EOS\space}%
\providecommand \EOS [0]{\spacefactor3000\relax}%
\providecommand \BibitemShut  [1]{\csname bibitem#1\endcsname}%
\let\auto@bib@innerbib\@empty
\bibitem [{\citenamefont {Novoselov}\ \emph {et~al.}(2004)\citenamefont
  {Novoselov}, \citenamefont {Geim}, \citenamefont {Morozov}, \citenamefont
  {Jiang}, \citenamefont {Zhang}, \citenamefont {Dubonos}, \citenamefont
  {Grigorieva},\ and\ \citenamefont {Firsov}}]{Novoselov_Electric_2004}%
  \BibitemOpen
  \bibfield  {author} {\bibinfo {author} {\bibfnamefont {K.}~\bibnamefont
  {Novoselov}}, \bibinfo {author} {\bibfnamefont {A.}~\bibnamefont {Geim}},
  \bibinfo {author} {\bibfnamefont {S.}~\bibnamefont {Morozov}}, \bibinfo
  {author} {\bibfnamefont {D.}~\bibnamefont {Jiang}}, \bibinfo {author}
  {\bibfnamefont {Y.}~\bibnamefont {Zhang}}, \bibinfo {author} {\bibfnamefont
  {S.}~\bibnamefont {Dubonos}}, \bibinfo {author} {\bibfnamefont
  {I.}~\bibnamefont {Grigorieva}}, \ and\ \bibinfo {author} {\bibfnamefont
  {A.}~\bibnamefont {Firsov}},\ }\href {\doibase 10.1126/science.1102896}
  {\bibfield  {journal} {\bibinfo  {journal} {Science}\ }\textbf {\bibinfo
  {volume} {306}},\ \bibinfo {pages} {666} (\bibinfo {year}
  {2004})}\BibitemShut {NoStop}%
\bibitem [{\citenamefont {Novoselov}\ \emph {et~al.}(2005)\citenamefont
  {Novoselov}, \citenamefont {Geim}, \citenamefont {Morozov}, \citenamefont
  {Jiang}, \citenamefont {Katsnelson}, \citenamefont {Grigorieva},
  \citenamefont {Dubonos},\ and\ \citenamefont {Firsov}}]{Novoselov_Two_2005}%
  \BibitemOpen
  \bibfield  {author} {\bibinfo {author} {\bibfnamefont {K.}~\bibnamefont
  {Novoselov}}, \bibinfo {author} {\bibfnamefont {A.}~\bibnamefont {Geim}},
  \bibinfo {author} {\bibfnamefont {S.}~\bibnamefont {Morozov}}, \bibinfo
  {author} {\bibfnamefont {D.}~\bibnamefont {Jiang}}, \bibinfo {author}
  {\bibfnamefont {M.}~\bibnamefont {Katsnelson}}, \bibinfo {author}
  {\bibfnamefont {I.}~\bibnamefont {Grigorieva}}, \bibinfo {author}
  {\bibfnamefont {S.}~\bibnamefont {Dubonos}}, \ and\ \bibinfo {author}
  {\bibfnamefont {A.}~\bibnamefont {Firsov}},\ }\href {\doibase
  10.1038/nature04233} {\bibfield  {journal} {\bibinfo  {journal} {Nature}\
  }\textbf {\bibinfo {volume} {438}},\ \bibinfo {pages} {197} (\bibinfo {year}
  {2005})}\BibitemShut {NoStop}%
\bibitem [{\citenamefont {Geim}\ and\ \citenamefont
  {Novoselov}(2007)}]{Geim_The_2007}%
  \BibitemOpen
  \bibfield  {author} {\bibinfo {author} {\bibfnamefont {A.}~\bibnamefont
  {Geim}}\ and\ \bibinfo {author} {\bibfnamefont {K.}~\bibnamefont
  {Novoselov}},\ }\href {\doibase 10.1038/nmat1849} {\bibfield  {journal}
  {\bibinfo  {journal} {Nature Materials}\ }\textbf {\bibinfo {volume} {6}},\
  \bibinfo {pages} {183} (\bibinfo {year} {2007})}\BibitemShut {NoStop}%
\bibitem [{\citenamefont {Geim}(2009)}]{Geim_Graphene_2009}%
  \BibitemOpen
  \bibfield  {author} {\bibinfo {author} {\bibfnamefont {A.}~\bibnamefont
  {Geim}},\ }\href {\doibase 10.1126/science.1158877} {\bibfield  {journal}
  {\bibinfo  {journal} {Science}\ }\textbf {\bibinfo {volume} {324}},\ \bibinfo
  {pages} {1530} (\bibinfo {year} {2009})}\BibitemShut {NoStop}%
\bibitem [{\citenamefont {Dean}\ \emph {et~al.}(2010)\citenamefont {Dean},
  \citenamefont {Young}, \citenamefont {Meric}, \citenamefont {Lee},
  \citenamefont {Wang}, \citenamefont {Sorgenfrei}, \citenamefont {Watanabe},
  \citenamefont {Taniguchi}, \citenamefont {Kim}, \citenamefont {Shepard},\
  and\ \citenamefont {Hone}}]{Dean_Boron_2010}%
  \BibitemOpen
  \bibfield  {author} {\bibinfo {author} {\bibfnamefont {C.}~\bibnamefont
  {Dean}}, \bibinfo {author} {\bibfnamefont {A.}~\bibnamefont {Young}},
  \bibinfo {author} {\bibfnamefont {I.}~\bibnamefont {Meric}}, \bibinfo
  {author} {\bibfnamefont {C.}~\bibnamefont {Lee}}, \bibinfo {author}
  {\bibfnamefont {L.}~\bibnamefont {Wang}}, \bibinfo {author} {\bibfnamefont
  {S.}~\bibnamefont {Sorgenfrei}}, \bibinfo {author} {\bibfnamefont
  {K.}~\bibnamefont {Watanabe}}, \bibinfo {author} {\bibfnamefont
  {T.}~\bibnamefont {Taniguchi}}, \bibinfo {author} {\bibfnamefont
  {P.}~\bibnamefont {Kim}}, \bibinfo {author} {\bibfnamefont {K.}~\bibnamefont
  {Shepard}}, \ and\ \bibinfo {author} {\bibfnamefont {J.}~\bibnamefont
  {Hone}},\ }\href {\doibase 10.1038/nnano.2010.172} {\bibfield  {journal}
  {\bibinfo  {journal} {Nature Nanotechnology}\ }\textbf {\bibinfo {volume}
  {5}},\ \bibinfo {pages} {722} (\bibinfo {year} {2010})}\BibitemShut {NoStop}%
\bibitem [{\citenamefont {Li}\ \emph {et~al.}(2009)\citenamefont {Li},
  \citenamefont {Zhu}, \citenamefont {Cai}, \citenamefont {Borysiak},
  \citenamefont {Han}, \citenamefont {Chen}, \citenamefont {Piner},
  \citenamefont {Colombo},\ and\ \citenamefont {Ruoff}}]{Li_Transfer_2009}%
  \BibitemOpen
  \bibfield  {author} {\bibinfo {author} {\bibfnamefont {X.}~\bibnamefont
  {Li}}, \bibinfo {author} {\bibfnamefont {Y.}~\bibnamefont {Zhu}}, \bibinfo
  {author} {\bibfnamefont {W.}~\bibnamefont {Cai}}, \bibinfo {author}
  {\bibfnamefont {M.}~\bibnamefont {Borysiak}}, \bibinfo {author}
  {\bibfnamefont {B.}~\bibnamefont {Han}}, \bibinfo {author} {\bibfnamefont
  {D.}~\bibnamefont {Chen}}, \bibinfo {author} {\bibfnamefont {R.~D.}\
  \bibnamefont {Piner}}, \bibinfo {author} {\bibfnamefont {L.}~\bibnamefont
  {Colombo}}, \ and\ \bibinfo {author} {\bibfnamefont {R.~S.}\ \bibnamefont
  {Ruoff}},\ }\href {\doibase 10.1021/nl902623y} {\bibfield  {journal}
  {\bibinfo  {journal} {Nano Letters}\ }\textbf {\bibinfo {volume} {9}},\
  \bibinfo {pages} {4359} (\bibinfo {year} {2009})}\BibitemShut {NoStop}%
\bibitem [{\citenamefont {Reina}\ \emph {et~al.}(2008)\citenamefont {Reina},
  \citenamefont {Son}, \citenamefont {Jiao}, \citenamefont {Fan}, \citenamefont
  {Dresselhaus}, \citenamefont {Liu},\ and\ \citenamefont
  {Kong}}]{Reina_Transferring_2008}%
  \BibitemOpen
  \bibfield  {author} {\bibinfo {author} {\bibfnamefont {A.}~\bibnamefont
  {Reina}}, \bibinfo {author} {\bibfnamefont {H.}~\bibnamefont {Son}}, \bibinfo
  {author} {\bibfnamefont {L.}~\bibnamefont {Jiao}}, \bibinfo {author}
  {\bibfnamefont {B.}~\bibnamefont {Fan}}, \bibinfo {author} {\bibfnamefont
  {M.~S.}\ \bibnamefont {Dresselhaus}}, \bibinfo {author} {\bibfnamefont
  {Z.}~\bibnamefont {Liu}}, \ and\ \bibinfo {author} {\bibfnamefont
  {J.}~\bibnamefont {Kong}},\ }\href {\doibase 10.1021/jp807380s} {\bibfield
  {journal} {\bibinfo  {journal} {Journal of Physical Chemistry C}\ }\textbf
  {\bibinfo {volume} {112}},\ \bibinfo {pages} {17741} (\bibinfo {year}
  {2008})}\BibitemShut {NoStop}%
\bibitem [{\citenamefont {Reina}\ \emph {et~al.}(2009)\citenamefont {Reina},
  \citenamefont {Jia}, \citenamefont {Ho}, \citenamefont {Nezich},
  \citenamefont {Son}, \citenamefont {Bulovic}, \citenamefont {Dresselhaus},\
  and\ \citenamefont {Kong}}]{Reina_Large_2009}%
  \BibitemOpen
  \bibfield  {author} {\bibinfo {author} {\bibfnamefont {A.}~\bibnamefont
  {Reina}}, \bibinfo {author} {\bibfnamefont {X.}~\bibnamefont {Jia}}, \bibinfo
  {author} {\bibfnamefont {J.}~\bibnamefont {Ho}}, \bibinfo {author}
  {\bibfnamefont {D.}~\bibnamefont {Nezich}}, \bibinfo {author} {\bibfnamefont
  {H.}~\bibnamefont {Son}}, \bibinfo {author} {\bibfnamefont {V.}~\bibnamefont
  {Bulovic}}, \bibinfo {author} {\bibfnamefont {M.~S.}\ \bibnamefont
  {Dresselhaus}}, \ and\ \bibinfo {author} {\bibfnamefont {J.}~\bibnamefont
  {Kong}},\ }\href {\doibase 10.1021/nl801827v} {\bibfield  {journal} {\bibinfo
   {journal} {Nano Letters}\ }\textbf {\bibinfo {volume} {9}},\ \bibinfo
  {pages} {30} (\bibinfo {year} {2009})}\BibitemShut {NoStop}%
\bibitem [{\citenamefont {Pirkle}\ \emph {et~al.}(2011)\citenamefont {Pirkle},
  \citenamefont {Chan}, \citenamefont {Venugopal}, \citenamefont {Hinojos},
  \citenamefont {Magnuson}, \citenamefont {S}, \citenamefont {Colombo},
  \citenamefont {Vogel}, \citenamefont {Ruoff},\ and\ \citenamefont
  {Wallace}}]{Pirkle_The_2011}%
  \BibitemOpen
  \bibfield  {author} {\bibinfo {author} {\bibfnamefont {A.}~\bibnamefont
  {Pirkle}}, \bibinfo {author} {\bibfnamefont {J.}~\bibnamefont {Chan}},
  \bibinfo {author} {\bibfnamefont {A.}~\bibnamefont {Venugopal}}, \bibinfo
  {author} {\bibfnamefont {D.}~\bibnamefont {Hinojos}}, \bibinfo {author}
  {\bibfnamefont {C.}~\bibnamefont {Magnuson}}, \bibinfo {author}
  {\bibfnamefont {M.}~\bibnamefont {S}}, \bibinfo {author} {\bibfnamefont
  {L.}~\bibnamefont {Colombo}}, \bibinfo {author} {\bibfnamefont
  {E.}~\bibnamefont {Vogel}}, \bibinfo {author} {\bibfnamefont
  {R.}~\bibnamefont {Ruoff}}, \ and\ \bibinfo {author} {\bibfnamefont
  {R.}~\bibnamefont {Wallace}},\ }\href {\doibase 10.1063/1.3643444} {\bibfield
   {journal} {\bibinfo  {journal} {Applied Phyics Letters}\ }\textbf {\bibinfo
  {volume} {99}},\ \bibinfo {pages} {122108} (\bibinfo {year}
  {2011})}\BibitemShut {NoStop}%
\bibitem [{\citenamefont {Lin}\ \emph {et~al.}(2012)\citenamefont {Lin},
  \citenamefont {Lu}, \citenamefont {Yeh}, \citenamefont {Jin}, \citenamefont
  {Suenaga},\ and\ \citenamefont {Chiu}}]{Lin_Graphene_2012}%
  \BibitemOpen
  \bibfield  {author} {\bibinfo {author} {\bibfnamefont {Y.}~\bibnamefont
  {Lin}}, \bibinfo {author} {\bibfnamefont {C.}~\bibnamefont {Lu}}, \bibinfo
  {author} {\bibfnamefont {C.}~\bibnamefont {Yeh}}, \bibinfo {author}
  {\bibfnamefont {C.}~\bibnamefont {Jin}}, \bibinfo {author} {\bibfnamefont
  {K.}~\bibnamefont {Suenaga}}, \ and\ \bibinfo {author} {\bibfnamefont
  {P.}~\bibnamefont {Chiu}},\ }\href {\doibase 10.1021/nl203733r} {\bibfield
  {journal} {\bibinfo  {journal} {Nano Letters}\ }\textbf {\bibinfo {volume}
  {12}},\ \bibinfo {pages} {414} (\bibinfo {year} {2012})}\BibitemShut
  {NoStop}%
\bibitem [{\citenamefont {Cheng}\ \emph {et~al.}(2011)\citenamefont {Cheng},
  \citenamefont {Zhou}, \citenamefont {Wang}, \citenamefont {Li}, \citenamefont
  {Wang},\ and\ \citenamefont {Fang}}]{Cheng_Toward_2011}%
  \BibitemOpen
  \bibfield  {author} {\bibinfo {author} {\bibfnamefont {Z.}~\bibnamefont
  {Cheng}}, \bibinfo {author} {\bibfnamefont {Q.}~\bibnamefont {Zhou}},
  \bibinfo {author} {\bibfnamefont {C.}~\bibnamefont {Wang}}, \bibinfo {author}
  {\bibfnamefont {Q.}~\bibnamefont {Li}}, \bibinfo {author} {\bibfnamefont
  {C.}~\bibnamefont {Wang}}, \ and\ \bibinfo {author} {\bibfnamefont
  {Y.}~\bibnamefont {Fang}},\ }\href {\doibase 10.1021/nl103977d} {\bibfield
  {journal} {\bibinfo  {journal} {Nano Letters}\ }\textbf {\bibinfo {volume}
  {11}},\ \bibinfo {pages} {767} (\bibinfo {year} {2011})}\BibitemShut
  {NoStop}%
\bibitem [{\citenamefont {Arcella}, \citenamefont {Troglia},\ and\
  \citenamefont {Ghielmi}(2005)}]{Arcella_Hyflon_2005}%
  \BibitemOpen
  \bibfield  {author} {\bibinfo {author} {\bibfnamefont {V.}~\bibnamefont
  {Arcella}}, \bibinfo {author} {\bibfnamefont {C.}~\bibnamefont {Troglia}}, \
  and\ \bibinfo {author} {\bibfnamefont {A.}~\bibnamefont {Ghielmi}},\ }\href
  {\doibase 10.1021/ie058008a} {\bibfield  {journal} {\bibinfo  {journal}
  {Industrial \& Engineering Chemistry Research}\ }\textbf {\bibinfo {volume}
  {44}},\ \bibinfo {pages} {7646} (\bibinfo {year} {2005})}\BibitemShut
  {NoStop}%
\bibitem [{\citenamefont {Merlo}\ \emph {et~al.}(2007)\citenamefont {Merlo},
  \citenamefont {Ghielmi}, \citenamefont {Cirillo},\ and\ \citenamefont
  {Gebert}}]{Merlo_Membrane_2007}%
  \BibitemOpen
  \bibfield  {author} {\bibinfo {author} {\bibfnamefont {L.}~\bibnamefont
  {Merlo}}, \bibinfo {author} {\bibfnamefont {A.}~\bibnamefont {Ghielmi}},
  \bibinfo {author} {\bibfnamefont {L.}~\bibnamefont {Cirillo}}, \ and\
  \bibinfo {author} {\bibfnamefont {M.}~\bibnamefont {Gebert}},\ }\href
  {\doibase 10.1080/01496390701558334} {\bibfield  {journal} {\bibinfo
  {journal} {Separation Science and Technology}\ }\textbf {\bibinfo {volume}
  {42}},\ \bibinfo {pages} {2891} (\bibinfo {year} {2007})}\BibitemShut
  {NoStop}%
\bibitem [{\citenamefont {Zhang}\ \emph {et~al.}(2012)\citenamefont {Zhang},
  \citenamefont {Chae}, \citenamefont {Hendren}, \citenamefont {Park},\ and\
  \citenamefont {Wiesner}}]{Zhang_Recent_2012}%
  \BibitemOpen
  \bibfield  {author} {\bibinfo {author} {\bibfnamefont {L.}~\bibnamefont
  {Zhang}}, \bibinfo {author} {\bibfnamefont {S.}~\bibnamefont {Chae}},
  \bibinfo {author} {\bibfnamefont {Z.}~\bibnamefont {Hendren}}, \bibinfo
  {author} {\bibfnamefont {J.}~\bibnamefont {Park}}, \ and\ \bibinfo {author}
  {\bibfnamefont {M.~R.}\ \bibnamefont {Wiesner}},\ }\href {\doibase
  10.1016/j.cej.2012.07.103} {\bibfield  {journal} {\bibinfo  {journal}
  {Chemical Engineering Journal}\ }\textbf {\bibinfo {volume} {204}},\ \bibinfo
  {pages} {87} (\bibinfo {year} {2012})}\BibitemShut {NoStop}%
\bibitem [{\citenamefont {Mattevi}\ \emph {et~al.}(2012)\citenamefont
  {Mattevi}, \citenamefont {Coll{\'e}aux}, \citenamefont {Kim}, \citenamefont
  {Lin}, \citenamefont {Park}, \citenamefont {Chhowalla},\ and\ \citenamefont
  {Anthopoulos}}]{Mattevi_Solution_2012}%
  \BibitemOpen
  \bibfield  {author} {\bibinfo {author} {\bibfnamefont {C.}~\bibnamefont
  {Mattevi}}, \bibinfo {author} {\bibfnamefont {F.}~\bibnamefont
  {Coll{\'e}aux}}, \bibinfo {author} {\bibfnamefont {H.}~\bibnamefont {Kim}},
  \bibinfo {author} {\bibfnamefont {Y.}~\bibnamefont {Lin}}, \bibinfo {author}
  {\bibfnamefont {K.~T.}\ \bibnamefont {Park}}, \bibinfo {author}
  {\bibfnamefont {M.}~\bibnamefont {Chhowalla}}, \ and\ \bibinfo {author}
  {\bibfnamefont {T.~D.}\ \bibnamefont {Anthopoulos}},\ }\href {\doibase
  10.1088/0957-4484/23/34/344017} {\bibfield  {journal} {\bibinfo  {journal}
  {Nanotechnology}\ }\textbf {\bibinfo {volume} {23}},\ \bibinfo {pages}
  {344017} (\bibinfo {year} {2012})}\BibitemShut {NoStop}%
\bibitem [{\citenamefont {Huang}\ \emph {et~al.}(2015)\citenamefont {Huang},
  \citenamefont {Jnawali}, \citenamefont {Hsu}, \citenamefont {Dhingra},
  \citenamefont {Lee}, \citenamefont {Ryu}, \citenamefont {Bi}, \citenamefont
  {Ghahari}, \citenamefont {Ravichandran}, \citenamefont {Chen}, \citenamefont
  {Kim}, \citenamefont {Eom}, \citenamefont {{D'Urso}}, \citenamefont {Irvin},\
  and\ \citenamefont {Levy}}]{Huang_Electric_2015}%
  \BibitemOpen
  \bibfield  {author} {\bibinfo {author} {\bibfnamefont {M.}~\bibnamefont
  {Huang}}, \bibinfo {author} {\bibfnamefont {G.}~\bibnamefont {Jnawali}},
  \bibinfo {author} {\bibfnamefont {J.}~\bibnamefont {Hsu}}, \bibinfo {author}
  {\bibfnamefont {S.}~\bibnamefont {Dhingra}}, \bibinfo {author} {\bibfnamefont
  {H.}~\bibnamefont {Lee}}, \bibinfo {author} {\bibfnamefont {S.}~\bibnamefont
  {Ryu}}, \bibinfo {author} {\bibfnamefont {F.}~\bibnamefont {Bi}}, \bibinfo
  {author} {\bibfnamefont {F.}~\bibnamefont {Ghahari}}, \bibinfo {author}
  {\bibfnamefont {J.}~\bibnamefont {Ravichandran}}, \bibinfo {author}
  {\bibfnamefont {L.}~\bibnamefont {Chen}}, \bibinfo {author} {\bibfnamefont
  {P.}~\bibnamefont {Kim}}, \bibinfo {author} {\bibfnamefont {C.}~\bibnamefont
  {Eom}}, \bibinfo {author} {\bibfnamefont {B.}~\bibnamefont {{D'Urso}}},
  \bibinfo {author} {\bibfnamefont {P.}~\bibnamefont {Irvin}}, \ and\ \bibinfo
  {author} {\bibfnamefont {J.}~\bibnamefont {Levy}},\ }\href {\doibase
  10.1063/1.4916098} {\bibfield  {journal} {\bibinfo  {journal} {APL
  Matererials}\ }\textbf {\bibinfo {volume} {3}},\ \bibinfo {pages} {062502}
  (\bibinfo {year} {2015})}\BibitemShut {NoStop}%
\bibitem [{\citenamefont {Ohtomo}\ and\ \citenamefont
  {Hwang}(2004)}]{Ohtomo_A_2004}%
  \BibitemOpen
  \bibfield  {author} {\bibinfo {author} {\bibfnamefont {A.}~\bibnamefont
  {Ohtomo}}\ and\ \bibinfo {author} {\bibfnamefont {H.}~\bibnamefont {Hwang}},\
  }\href {\doibase 10.1038/nature02308} {\bibfield  {journal} {\bibinfo
  {journal} {Nature}\ }\textbf {\bibinfo {volume} {427}},\ \bibinfo {pages}
  {423} (\bibinfo {year} {2004})}\BibitemShut {NoStop}%
\bibitem [{\citenamefont {Thiel}\ \emph {et~al.}(2006)\citenamefont {Thiel},
  \citenamefont {Hammerl}, \citenamefont {Schmehl}, \citenamefont {Schneider},\
  and\ \citenamefont {Mannhart}}]{Thiel_Tunable_2006}%
  \BibitemOpen
  \bibfield  {author} {\bibinfo {author} {\bibfnamefont {S.}~\bibnamefont
  {Thiel}}, \bibinfo {author} {\bibfnamefont {G.}~\bibnamefont {Hammerl}},
  \bibinfo {author} {\bibfnamefont {A.}~\bibnamefont {Schmehl}}, \bibinfo
  {author} {\bibfnamefont {C.}~\bibnamefont {Schneider}}, \ and\ \bibinfo
  {author} {\bibfnamefont {J.}~\bibnamefont {Mannhart}},\ }\href {\doibase
  10.1126/science.1131091} {\bibfield  {journal} {\bibinfo  {journal}
  {Science}\ }\textbf {\bibinfo {volume} {313}},\ \bibinfo {pages} {1942}
  (\bibinfo {year} {2006})}\BibitemShut {NoStop}%
\bibitem [{\citenamefont {Brinkman}\ \emph {et~al.}(2007)\citenamefont
  {Brinkman}, \citenamefont {Huijben}, \citenamefont {van Zalk}, \citenamefont
  {Huijben}, \citenamefont {Zeitler}, \citenamefont {Maan}, \citenamefont
  {van~der Wiel}, \citenamefont {Rijnders}, \citenamefont {Blank},\ and\
  \citenamefont {Hilgenkamp}}]{Brinkman_Magnetic_2007}%
  \BibitemOpen
  \bibfield  {author} {\bibinfo {author} {\bibfnamefont {A.}~\bibnamefont
  {Brinkman}}, \bibinfo {author} {\bibfnamefont {M.}~\bibnamefont {Huijben}},
  \bibinfo {author} {\bibfnamefont {M.}~\bibnamefont {van Zalk}}, \bibinfo
  {author} {\bibfnamefont {J.}~\bibnamefont {Huijben}}, \bibinfo {author}
  {\bibfnamefont {U.}~\bibnamefont {Zeitler}}, \bibinfo {author} {\bibfnamefont
  {J.}~\bibnamefont {Maan}}, \bibinfo {author} {\bibfnamefont {W.}~\bibnamefont
  {van~der Wiel}}, \bibinfo {author} {\bibfnamefont {G.}~\bibnamefont
  {Rijnders}}, \bibinfo {author} {\bibfnamefont {D.}~\bibnamefont {Blank}}, \
  and\ \bibinfo {author} {\bibfnamefont {H.}~\bibnamefont {Hilgenkamp}},\
  }\href {\doibase 10.1038/nmat1931} {\bibfield  {journal} {\bibinfo  {journal}
  {Nature Materials}\ }\textbf {\bibinfo {volume} {6}},\ \bibinfo {pages} {493}
  (\bibinfo {year} {2007})}\BibitemShut {NoStop}%
\bibitem [{\citenamefont {Li}\ \emph {et~al.}(2011)\citenamefont {Li},
  \citenamefont {Richter}, \citenamefont {Mannhart},\ and\ \citenamefont
  {Ashoori}}]{Li_Coexistence_2011}%
  \BibitemOpen
  \bibfield  {author} {\bibinfo {author} {\bibfnamefont {L.}~\bibnamefont
  {Li}}, \bibinfo {author} {\bibfnamefont {C.}~\bibnamefont {Richter}},
  \bibinfo {author} {\bibfnamefont {J.}~\bibnamefont {Mannhart}}, \ and\
  \bibinfo {author} {\bibfnamefont {R.}~\bibnamefont {Ashoori}},\ }\href
  {\doibase 10.1038/nphys2080} {\bibfield  {journal} {\bibinfo  {journal}
  {Nature Physics}\ }\textbf {\bibinfo {volume} {7}},\ \bibinfo {pages} {762}
  (\bibinfo {year} {2011})}\BibitemShut {NoStop}%
\bibitem [{\citenamefont {Bi}\ \emph {et~al.}(2014)\citenamefont {Bi},
  \citenamefont {Huang}, \citenamefont {Ryu}, \citenamefont {Lee},
  \citenamefont {Bark}, \citenamefont {Eom}, \citenamefont {Irvin},\ and\
  \citenamefont {Levy}}]{Bi_Room_2014}%
  \BibitemOpen
  \bibfield  {author} {\bibinfo {author} {\bibfnamefont {F.}~\bibnamefont
  {Bi}}, \bibinfo {author} {\bibfnamefont {M.}~\bibnamefont {Huang}}, \bibinfo
  {author} {\bibfnamefont {S.}~\bibnamefont {Ryu}}, \bibinfo {author}
  {\bibfnamefont {H.}~\bibnamefont {Lee}}, \bibinfo {author} {\bibfnamefont
  {C.}~\bibnamefont {Bark}}, \bibinfo {author} {\bibfnamefont {C.}~\bibnamefont
  {Eom}}, \bibinfo {author} {\bibfnamefont {P.}~\bibnamefont {Irvin}}, \ and\
  \bibinfo {author} {\bibfnamefont {J.}~\bibnamefont {Levy}},\ }\href {\doibase
  10.1038/ncomms6019} {\bibfield  {journal} {\bibinfo  {journal} {Nature
  Communication}\ }\textbf {\bibinfo {volume} {5}},\ \bibinfo {pages} {5019}
  (\bibinfo {year} {2014})}\BibitemShut {NoStop}%
\bibitem [{\citenamefont {Reyren}\ \emph {et~al.}(2007)\citenamefont {Reyren},
  \citenamefont {Thiel}, \citenamefont {Caviglia}, \citenamefont {Kourkoutis},
  \citenamefont {Hammerl}, \citenamefont {Richter}, \citenamefont {Schneider},
  \citenamefont {Kopp}, \citenamefont {R{\"u}etschi}, \citenamefont {Jaccard},
  \citenamefont {Gabay}, \citenamefont {Muller}, \citenamefont {Triscone},\
  and\ \citenamefont {Mannhart}}]{Reyren_Superconducting_2007}%
  \BibitemOpen
  \bibfield  {author} {\bibinfo {author} {\bibfnamefont {N.}~\bibnamefont
  {Reyren}}, \bibinfo {author} {\bibfnamefont {S.}~\bibnamefont {Thiel}},
  \bibinfo {author} {\bibfnamefont {A.}~\bibnamefont {Caviglia}}, \bibinfo
  {author} {\bibfnamefont {L.}~\bibnamefont {Kourkoutis}}, \bibinfo {author}
  {\bibfnamefont {G.}~\bibnamefont {Hammerl}}, \bibinfo {author} {\bibfnamefont
  {C.}~\bibnamefont {Richter}}, \bibinfo {author} {\bibfnamefont
  {C.}~\bibnamefont {Schneider}}, \bibinfo {author} {\bibfnamefont
  {T.}~\bibnamefont {Kopp}}, \bibinfo {author} {\bibfnamefont {A.~S.}\
  \bibnamefont {R{\"u}etschi}}, \bibinfo {author} {\bibfnamefont
  {D.}~\bibnamefont {Jaccard}}, \bibinfo {author} {\bibfnamefont
  {M.}~\bibnamefont {Gabay}}, \bibinfo {author} {\bibfnamefont
  {D.}~\bibnamefont {Muller}}, \bibinfo {author} {\bibfnamefont {J.~M.}\
  \bibnamefont {Triscone}}, \ and\ \bibinfo {author} {\bibfnamefont
  {J.}~\bibnamefont {Mannhart}},\ }\href {\doibase 10.1126/science.1146006}
  {\bibfield  {journal} {\bibinfo  {journal} {Science}\ }\textbf {\bibinfo
  {volume} {317}},\ \bibinfo {pages} {1196} (\bibinfo {year}
  {2007})}\BibitemShut {NoStop}%
\bibitem [{\citenamefont {Cheng}\ \emph {et~al.}(2015)\citenamefont {Cheng},
  \citenamefont {Tomczyk}, \citenamefont {Lu}, \citenamefont {Veazey},
  \citenamefont {Huang}, \citenamefont {Irvin}, \citenamefont {Ryu},
  \citenamefont {Lee}, \citenamefont {Eom}, \citenamefont {Hellberg},\ and\
  \citenamefont {Levy}}]{Cheng_Electron_2015}%
  \BibitemOpen
  \bibfield  {author} {\bibinfo {author} {\bibfnamefont {G.}~\bibnamefont
  {Cheng}}, \bibinfo {author} {\bibfnamefont {M.}~\bibnamefont {Tomczyk}},
  \bibinfo {author} {\bibfnamefont {S.}~\bibnamefont {Lu}}, \bibinfo {author}
  {\bibfnamefont {J.~P.}\ \bibnamefont {Veazey}}, \bibinfo {author}
  {\bibfnamefont {M.}~\bibnamefont {Huang}}, \bibinfo {author} {\bibfnamefont
  {P.}~\bibnamefont {Irvin}}, \bibinfo {author} {\bibfnamefont
  {S.}~\bibnamefont {Ryu}}, \bibinfo {author} {\bibfnamefont {H.}~\bibnamefont
  {Lee}}, \bibinfo {author} {\bibfnamefont {C.}~\bibnamefont {Eom}}, \bibinfo
  {author} {\bibfnamefont {S.~C.}\ \bibnamefont {Hellberg}}, \ and\ \bibinfo
  {author} {\bibfnamefont {J.}~\bibnamefont {Levy}},\ }\href {\doibase
  10.1038/nature14398} {\bibfield  {journal} {\bibinfo  {journal} {Nature}\
  }\textbf {\bibinfo {volume} {521}},\ \bibinfo {pages} {196} (\bibinfo {year}
  {2015})}\BibitemShut {NoStop}%
\bibitem [{\citenamefont {Cen}\ \emph {et~al.}(2008)\citenamefont {Cen},
  \citenamefont {Thiel}, \citenamefont {Hammerl}, \citenamefont {Schneider},
  \citenamefont {Andersen}, \citenamefont {Hellberg}, \citenamefont
  {Mannhart},\ and\ \citenamefont {Levy}}]{Cen_Nanoscale_2008}%
  \BibitemOpen
  \bibfield  {author} {\bibinfo {author} {\bibfnamefont {C.}~\bibnamefont
  {Cen}}, \bibinfo {author} {\bibfnamefont {S.}~\bibnamefont {Thiel}}, \bibinfo
  {author} {\bibfnamefont {G.}~\bibnamefont {Hammerl}}, \bibinfo {author}
  {\bibfnamefont {C.}~\bibnamefont {Schneider}}, \bibinfo {author}
  {\bibfnamefont {K.}~\bibnamefont {Andersen}}, \bibinfo {author}
  {\bibfnamefont {C.}~\bibnamefont {Hellberg}}, \bibinfo {author}
  {\bibfnamefont {J.}~\bibnamefont {Mannhart}}, \ and\ \bibinfo {author}
  {\bibfnamefont {J.}~\bibnamefont {Levy}},\ }\href {\doibase 10.1038/nmat2136}
  {\bibfield  {journal} {\bibinfo  {journal} {Nature Materials}\ }\textbf
  {\bibinfo {volume} {7}},\ \bibinfo {pages} {298} (\bibinfo {year}
  {2008})}\BibitemShut {NoStop}%
\bibitem [{\citenamefont {Cen}\ \emph {et~al.}(2009)\citenamefont {Cen},
  \citenamefont {Thiel}, \citenamefont {Mannhart},\ and\ \citenamefont
  {Levy}}]{Cen_Oxide_2009}%
  \BibitemOpen
  \bibfield  {author} {\bibinfo {author} {\bibfnamefont {C.}~\bibnamefont
  {Cen}}, \bibinfo {author} {\bibfnamefont {S.}~\bibnamefont {Thiel}}, \bibinfo
  {author} {\bibfnamefont {J.}~\bibnamefont {Mannhart}}, \ and\ \bibinfo
  {author} {\bibfnamefont {J.}~\bibnamefont {Levy}},\ }\href {\doibase
  10.1126/science.1168294} {\bibfield  {journal} {\bibinfo  {journal}
  {Science}\ }\textbf {\bibinfo {volume} {323}},\ \bibinfo {pages} {1026}
  (\bibinfo {year} {2009})}\BibitemShut {NoStop}%
\bibitem [{\citenamefont {Jnawali}\ \emph {et~al.}(2016)\citenamefont
  {Jnawali}, \citenamefont {Huang}, \citenamefont {Hsu}, \citenamefont {Lee},
  \citenamefont {Irvin}, \citenamefont {Eom}, \citenamefont {D'Urso},\ and\
  \citenamefont {Levy}}]{jnawali2016room}%
  \BibitemOpen
  \bibfield  {author} {\bibinfo {author} {\bibfnamefont {G.}~\bibnamefont
  {Jnawali}}, \bibinfo {author} {\bibfnamefont {M.}~\bibnamefont {Huang}},
  \bibinfo {author} {\bibfnamefont {J.-F.}\ \bibnamefont {Hsu}}, \bibinfo
  {author} {\bibfnamefont {H.}~\bibnamefont {Lee}}, \bibinfo {author}
  {\bibfnamefont {P.}~\bibnamefont {Irvin}}, \bibinfo {author} {\bibfnamefont
  {C.-B.}\ \bibnamefont {Eom}}, \bibinfo {author} {\bibfnamefont
  {B.}~\bibnamefont {D'Urso}}, \ and\ \bibinfo {author} {\bibfnamefont
  {J.}~\bibnamefont {Levy}},\ }\href@noop {} {\bibfield  {journal} {\bibinfo
  {journal} {arXiv preprint arXiv:1602.03128}\ } (\bibinfo {year}
  {2016})}\BibitemShut {NoStop}%
\bibitem [{\citenamefont {Dhingra}\ \emph {et~al.}(2014)\citenamefont
  {Dhingra}, \citenamefont {Hsu}, \citenamefont {Vlassiouk},\ and\
  \citenamefont {{D'Urso}}}]{Dhingra_Chemical_2014}%
  \BibitemOpen
  \bibfield  {author} {\bibinfo {author} {\bibfnamefont {S.}~\bibnamefont
  {Dhingra}}, \bibinfo {author} {\bibfnamefont {J.}~\bibnamefont {Hsu}},
  \bibinfo {author} {\bibfnamefont {I.}~\bibnamefont {Vlassiouk}}, \ and\
  \bibinfo {author} {\bibfnamefont {B.}~\bibnamefont {{D'Urso}}},\ }\href
  {\doibase 10.1016/j.carbon.2013.12.014} {\bibfield  {journal} {\bibinfo
  {journal} {Carbon}\ }\textbf {\bibinfo {volume} {69}},\ \bibinfo {pages}
  {188} (\bibinfo {year} {2014})}\BibitemShut {NoStop}%
\bibitem [{\citenamefont {Goossens}\ \emph {et~al.}(2012)\citenamefont
  {Goossens}, \citenamefont {Calado}, \citenamefont {Barreiro}, \citenamefont
  {Watanabe}, \citenamefont {Taniguchi},\ and\ \citenamefont
  {Vandersypen}}]{Goossens_Mechanical_2012}%
  \BibitemOpen
  \bibfield  {author} {\bibinfo {author} {\bibfnamefont {A.}~\bibnamefont
  {Goossens}}, \bibinfo {author} {\bibfnamefont {V.}~\bibnamefont {Calado}},
  \bibinfo {author} {\bibfnamefont {A.}~\bibnamefont {Barreiro}}, \bibinfo
  {author} {\bibfnamefont {K.}~\bibnamefont {Watanabe}}, \bibinfo {author}
  {\bibfnamefont {T.}~\bibnamefont {Taniguchi}}, \ and\ \bibinfo {author}
  {\bibfnamefont {L.}~\bibnamefont {Vandersypen}},\ }\href {\doibase
  10.1063/1.3685504} {\bibfield  {journal} {\bibinfo  {journal} {Applied Phyics
  Letters}\ }\textbf {\bibinfo {volume} {100}},\ \bibinfo {pages} {073110}
  (\bibinfo {year} {2012})}\BibitemShut {NoStop}%
\bibitem [{\citenamefont {Lindvall}, \citenamefont {Kalabukhov},\ and\
  \citenamefont {Yurgens}(2012)}]{Lindvall_Cleaning_2012}%
  \BibitemOpen
  \bibfield  {author} {\bibinfo {author} {\bibfnamefont {N.}~\bibnamefont
  {Lindvall}}, \bibinfo {author} {\bibfnamefont {A.}~\bibnamefont
  {Kalabukhov}}, \ and\ \bibinfo {author} {\bibfnamefont {A.}~\bibnamefont
  {Yurgens}},\ }\href {\doibase 10.1063/1.3695451} {\bibfield  {journal}
  {\bibinfo  {journal} {Journal of Applied Physics}\ }\textbf {\bibinfo
  {volume} {111}},\ \bibinfo {pages} {064904} (\bibinfo {year}
  {2012})}\BibitemShut {NoStop}%
\bibitem [{\citenamefont {Weaver}(1959)}]{Weaver_Dielectric_1959}%
  \BibitemOpen
  \bibfield  {author} {\bibinfo {author} {\bibfnamefont {H.}~\bibnamefont
  {Weaver}},\ }\href {\doibase 10.1016/0022-3697(59)90226-4} {\bibfield
  {journal} {\bibinfo  {journal} {Journal of Physics and Chemistry of Solids}\
  }\textbf {\bibinfo {volume} {11}},\ \bibinfo {pages} {274} (\bibinfo {year}
  {1959})}\BibitemShut {NoStop}%
\bibitem [{\citenamefont {Huang}\ \emph {et~al.}(2013)\citenamefont {Huang},
  \citenamefont {Bi}, \citenamefont {Ryu}, \citenamefont {Eom}, \citenamefont
  {Irvin},\ and\ \citenamefont {Levy}}]{Huang_Direct_2013}%
  \BibitemOpen
  \bibfield  {author} {\bibinfo {author} {\bibfnamefont {M.}~\bibnamefont
  {Huang}}, \bibinfo {author} {\bibfnamefont {F.}~\bibnamefont {Bi}}, \bibinfo
  {author} {\bibfnamefont {S.}~\bibnamefont {Ryu}}, \bibinfo {author}
  {\bibfnamefont {C.}~\bibnamefont {Eom}}, \bibinfo {author} {\bibfnamefont
  {P.}~\bibnamefont {Irvin}}, \ and\ \bibinfo {author} {\bibfnamefont
  {J.}~\bibnamefont {Levy}},\ }\href {\doibase 10.1063/1.4831855} {\bibfield
  {journal} {\bibinfo  {journal} {APL Matererials}\ }\textbf {\bibinfo {volume}
  {1}},\ \bibinfo {pages} {052110} (\bibinfo {year} {2013})}\BibitemShut
  {NoStop}%
\end{thebibliography}%

\newpage

\newpage

\newpage

\newpage

\end{document}